\documentclass[12pt]{article}

\usepackage[height=8.85in,width=6.45in]{geometry}

\usepackage[utf8]{inputenc}
\usepackage[T1]{fontenc}
\usepackage{amsmath}
\usepackage{amssymb}
\usepackage{mathtools}
\numberwithin{equation}{section}
\allowdisplaybreaks

\usepackage{slashed}
\usepackage{braket}
\usepackage[usenames,dvipsnames,svgnames,table]{xcolor}
\usepackage[colorlinks,citecolor=DarkGreen,linkcolor=FireBrick,urlcolor=FireBrick,linktocpage]{hyperref}
\usepackage{cite}
\usepackage{graphicx}
\usepackage{times}
\usepackage[scaled]{couriers}

\usepackage{bm}
\usepackage{subfig}

\usepackage{tikz,pgf}
\usepackage{tikz-cd}
\usetikzlibrary{shapes}
\usetikzlibrary{calc}
\usetikzlibrary{decorations.pathmorphing}
\usetikzlibrary{decorations.pathreplacing,shapes.misc}
\usetikzlibrary{positioning}
\usetikzlibrary{decorations.markings}
\tikzset{->-/.style={decoration={
  markings,
  mark=at position .5 with {\arrow{>}}},postaction={decorate}}}
\tikzset{-<-/.style={decoration={
  markings,
  mark=at position .5 with {\arrow{<}}},postaction={decorate}}}

\usepackage{xcolor}
  \definecolor{rblue}{RGB}{81, 49, 193}
  \definecolor{rorange}{RGB}{255, 147, 40}
  \definecolor{rgreen}{RGB}{176, 233, 0}

\newcommand{\p}{\partial}

\renewcommand{\tilde}{\widetilde}

\newcommand{\bRP}{\mathbb{RP}}
\newcommand{\sfR}{{\sf R}}

\DeclareMathOperator{\Hom}{Hom}

\newcommand{\calC}{\mathcal{C}}
\newcommand{\calD}{\mathcal{D}}
\newcommand{\calV}{\mathcal{V}}

\renewcommand{\hat}{\widehat}

\usepackage{mdframed}

\def\Hom{\mathop{\mathrm{Hom}}}
\def\Sq{\mathop{\mathrm{Sq}}\nolimits}
\def\bZ{\mathbb{Z}}

\begin{document}

\begin{center}

{\large \bfseries Pin TQFT and Grassmann integral}

\bigskip
\bigskip
\bigskip

Ryohei Kobayashi
\bigskip
\bigskip
\bigskip

\begin{tabular}{ll}
 Institute for Solid State Physics, \\
University of Tokyo, Kashiwa, Chiba 277-8583, Japan\\

\end{tabular}

\vskip 1cm

\end{center}

\noindent 
We discuss a recipe to produce a lattice construction of fermionic phases of matter on unoriented manifolds. This is performed by extending the construction of spin TQFT via the Grassmann integral proposed by Gaiotto and Kapustin, to the unoriented pin$_\pm$ case. As an application, we construct gapped boundaries for time-reversal-invariant Gu-Wen fermionic SPT phases. In addition, we provide a lattice definition of (1+1)d pin$_-$ invertible theory whose partition function is the Arf-Brown-Kervaire invariant, which generates the $\bZ_8$ classification of (1+1)d topological superconductors. We also compute the indicator formula of $\bZ_{16}$ valued time-reversal anomaly for (2+1)d pin$_+$ TQFT based on our construction.

\setcounter{tocdepth}{2}
\tableofcontents

\section{Introduction and summary}

The notion of fermionic topological phase of matter has
attracted great interest, since fermionic systems admit novel phases that have no counterpart in bosonic systems~\cite{Gaiotto:2015zta,Gu:2012ib, Wang2017Interacting, Witten2016Fermion, Metlitski2014, Cheng2018Classification, Gu2014Lattice, Guo:2018vij}.

On orientable spacetime, 
fermionic topological phases are thought to be described at long distances by spin Topological Quantum Field Theory (spin TQFT).
In~\cite{Gaiotto:2015zta}, the authors provided a recipe to construct a state sum definition of spin TQFT, by formulating the spin theory called 
the Gu-Wen Grassmann integral on an oriented spin $d$-manifold $M$, equipped with a $(d-2)$-form $\bZ_2$ symmetry, whose partition function has the form
\begin{equation}
z[M, \eta,\alpha]=\sigma(M, \alpha)(-1)^{\int_M\eta\cup\alpha},
\end{equation}
where $\alpha\in Z^{d-1}(M, \bZ_2)$ is a background $\mathbb{Z}_2$
gauge field of the $(d-2)$-form symmetry, and $\eta$ specifies a spin structure of $M$, which is related to a 2-cocycle $w_2$ representing the second Stiefel-Whitney class 
as $\delta \eta = w_2$.
$\sigma(M, \alpha)$ is written in terms of a certain path integral of Grassmann variables defined by giving a triangulation of $M$.
(In the following, when there is no confusion, 
we simply write $z[\eta, \alpha]$, $\sigma(\alpha)$, instead of 
$z[M,\eta,\alpha]$, $\sigma(M,\alpha)$, etc.)

By studying the effect of re-triangulations and gauge transformations, this theory is shown to have an anomaly characterized by $(-1)^{\int \Sq^2(\alpha)}$, where $\Sq^2(\alpha)$ is the Steenrod square defined as $\Sq^2(\alpha):=\alpha\cup_{d-3}\alpha$. Then, one can construct a spin theory fully invariant under the change of triangulation and gauge transformations,
by coupling the Grassmann integral with a non-spin theory $\widetilde{Z}[\alpha]$ called a ``shadow theory''~\cite{Bhardwaj2017Statesum, Ellison2019}, whose anomaly is again characterized by  $(-1)^{\int \Sq^2(\alpha)}$, and then gauging the $(d-2)$-form symmetry,
\begin{equation}
Z[\eta]=\sum_{\alpha}z[\eta, \alpha]\widetilde{Z}[\alpha].
\end{equation}

In contrast, 
it is sometimes useful to consider
a fermionic topological phase on an unoriented manifold~\cite{Kapustin:2014dxa, Witten:2016cio, Bhardwaj2017,  Shapourian2017, Turzillo2018}, 
when the system has a symmetry that reverses the orientation of spacetime.
In such a situation, the corresponding theory requires a pin structure, 
which encodes the orientation reversing symmetry. 
For instance, let us think of 
a (1+1)d topological superconductor in class BDI (characterized by time reversal symmetry with $T^2=1$), which follows a $\bZ_8$ classification~\cite{FidkowskiKitaev2011}. Cobordism theory~\cite{Kapustin:2014dxa, Freed:2016rqq, Yonekura:2018ufj} predicts that the $\bZ_8$ classification is diagnosed by computing the partition function of the corresponding TQFT on an unoriented surface $\bRP^2$ equipped with a pin$_-$ structure. As another example, the (3+1)d topological superconductor in class DIII (time reversal symmetry with $T^2=(-1)^F$) is known to be classified by $\bZ_{16}$~\cite{Metlitski2014, Fidkowski2014, Hsieh2016, Tachikawa2017more}. The $\bZ_{16}$ classification is detected by the partition function of the TQFT on $\bRP^4$, equipped with a pin$_+$ structure. In this context, it is important to ask how to formulate the pin$_\pm$ TQFT on a manifold which is not necessarily oriented.

In this paper, we propose a strategy to produce a lattice definition of 
pin$_\pm$ TQFT in general dimensions,
by extending the recipe in~\cite{Gaiotto:2015zta}.
Concretely, we obtain the extended Grassmann integral 
$\sigma(M, \alpha)$ on an unoriented $d$-manifold $M$. 
This is done
by modifying the definition of the Grassmann integral properly, in the vicinity of the orientation reversing wall in $M$, which flips the orientation as we go across the wall.
We will show that the effect of re-triangulation and gauge transformation is expressed as
\begin{equation}
\sigma(\tilde{M}, \tilde{\alpha})=(-1)^{\int_K (\Sq^2(\alpha)+(w_2+w_1^2)\cup\alpha)}\sigma(M,\alpha),
\end{equation}
where $\tilde M$ is the same manifold $M$ with a different triangulation,
$\tilde\alpha$ is a cocycle such that $[\alpha]=[\tilde\alpha] $ in cohomology,
and $K=M\times [0,1]$ such that the two boundaries are given by $M$ and $\tilde M$,
and finally $\alpha$ is extended to $K$ so that it restricts to $\alpha$ and $\tilde \alpha$ on the boundaries.

Then, we can define the pin$_-$ TQFT when $M$ admits a pin$_-$ structure, by coupling with a bosonic shadow theory $\widetilde{Z}_-[\alpha]$ which possesses an anomaly $(-1)^{\int \Sq^2(\alpha)}$,
\begin{equation}
Z_{\mathrm{pin}_-}[M, \eta]=\sum_{\alpha}\widetilde{Z}_-[\alpha]\sigma(M, \alpha)(-1)^{\int_M\eta\cup\alpha},
\end{equation}
where $\eta$ specifies a pin$_-$ structure that satisfies $\delta\eta=w_2+w_1^2$. We can also construct the pin$_+$ TQFT when $M$ admits a pin$_+$ structure, by coupling with a bosonic shadow theory $\widetilde{Z}_+[\alpha]$ with an anomaly $(-1)^{\int \Sq^2(\alpha)+w_1^2\cup\alpha}$,
\begin{equation}
Z_{\mathrm{pin}_+}[M, \eta]=\sum_{\alpha}\widetilde{Z}_+[\alpha]\sigma(M, \alpha)(-1)^{\int_M\eta\cup\alpha},
\end{equation}
where $\eta$ specifies a pin$_+$ structure that satisfies $\delta\eta=w_2$.

We have several applications of our construction of pin$_{\pm}$ TQFT based on the Grassmann integral. First, we construct the TQFT for a subclass of fermionic SPT phases known as pin$_{\pm}$ Gu-Wen $G$-SPT phases~\cite{Gu:2012ib,Gaiotto:2015zta}. We further show that pin$_{\pm}$ Gu-Wen SPT phases always admit a
gapped boundary, by explicitly constructing the Grassmann integral 
for the coupled bulk and boundary system on an unoriented manifold.
In addition, we propose a lattice definition of 2d pin$_-$ TQFT whose partition function is
the Arf-Brown-Kervaire (ABK) invariant~\cite{Brown1972, Debray2018, Turzillo2018}, which generates the $\bZ_8$ classification of (1+1)d topological superconductors. 
Finally, we discuss a way to compute the $\bZ_{16}$-valued (2+1)d pin$_+$ anomaly from the data of (2+1)d anomalous theory. 
Such a formula for the $\bZ_{16}$ anomaly (known as 
the indicator formula) has been conjectured in~\cite{Wang2017indicator}, and later proven in~\cite{Tachikawa2017more}. 
We compute the indicator formula when the anomalous theory is a pin$_+$ TQFT whose shadow theory in the bulk is given by the (3+1)d Walker-Wang model~\cite{WalkerWang2011}.
Our indicator formula is expressed in terms of the data of the shadow TQFT.

This paper is organized as follows. In Sec.~\ref{sec:GuWen}, we review the construction of the Grassmann integral for the oriented case, and describe
the spin TQFT for the Gu-Wen SPT phase.
In Sec.~\ref{sec:pinGuWen}, we construct an extended Gu-Wen integral for unoriented manifolds, and describe the Gu-Wen $G$-SPT phase based on the pin$_{\pm}$ structure.
In Sec.~\ref{sec:ABK}, we propose a lattice construction of the ABK invariant based on the Grassmann integral. In Sec.~\ref{sec:boundary}, we construct gapped boundary theories for the Gu-Wen pin$_{\pm}$ $G$-SPT phases. Finally, in Sec.~\ref{sec:anomaly}, we compute the indicator formula for $\bZ_{16}$-valued anomaly of (2+1)d pin$_{+}$ TQFT.

\section{Review: Grassmann integral and Gu-Wen spin SPT phases}
\label{sec:GuWen}
In this section, we first recall the construction of the Grassmann integral on an oriented spin $d$-manifold $M$ formulated in~\cite{Gaiotto:2015zta}. Next, we describe the spin TQFT for fermionic Gu-Wen $G$-SPT phases.

\subsection{Review of the Gu-Wen Grassmann integral for spin case}
We first endow $M$ with a triangulation. In addition, we take the
barycentric subdivision for the triangulation of $M$. Namely, each $d$-simplex in the initial triangulation of $M$ is subdivided into $(d+1)!$ simplices, whose vertices are barycenters of the subsets of vertices in the $d$-simplex. We further assign a local ordering to vertices of the barycentric subdivision, such that a vertex on the barycenter of $i$ vertices is labeled as $i$.

Each simplex can then be either a $+$ simplex or a $-$ simplex, depending on whether the ordering agrees with the orientation or not.
We assign a pair of Grassmann variables $\theta_e, \overline{\theta}_e$ on each $(d-1)$-simplex $e$ of $M$ such that $\alpha(e)=1$, we associate $\theta_e$ on one side of $e$ contained in one of $d$-simplices neighboring $e$ (which will be specified later), $\overline{\theta}_e$ on the other side.
Then, $\sigma(M, \alpha)$ is defined as
\begin{equation}
    \sigma(M, \alpha)=\int\prod_{e|\alpha(e)=1}d\theta_e d\overline{\theta}_e \prod_t u(t),
    \label{sigmadef}
\end{equation}
where $t$ denotes a $d$-simplex, and $u(t)$ is the product of Grassmann variables contained in $t$.
For instance, for $d=2$, $u(t)$ on $t=(012)$ is the product of
$\vartheta_{12}^{\alpha(12)}, \vartheta_{01}^{\alpha(01)}, \vartheta_{02}^{\alpha(02)}$. 
Here, $\vartheta$ denotes $\theta$ or $\overline{\theta}$ depending on the choice of the assigning rule, which will be discussed later. The order of Grassmann variables in $u(t)$ will also be defined shortly.
We note that $u(t)$ is ensured to be Grassmann-even when $\alpha$ is closed. 

\begin{figure}[htb]
\centering
\includegraphics{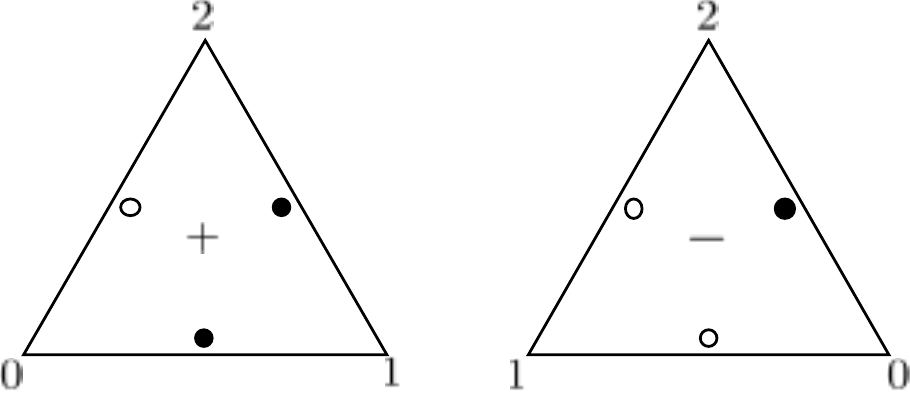}
\caption{Assignment of Grassmann variables on 1-simplices in the case of $d=2$. $\theta$ (resp.~$\overline{\theta}$) is represented as a black (resp.~white) dot.}
\label{fig:Grassmann}
\end{figure}

Due to the fermionic sign of Grassmann variables, $\sigma(\alpha)$ becomes a quadratic function, whose quadratic property depends on the order of Grassmann variables in $u(t)$. We will adopt the order used in Gaiotto-Kapustin \cite{Gaiotto:2015zta}, which is defined as follows. 
\begin{itemize}
\item
For $t=(01\dots d)$, we label a $(d-1)$-simplex $(01\dots\hat{i}\dots d)$ (i.e., a $(d-1)$-simplex given by omitting a vertex $i$) simply as $i$. 
\item Then, the order of $\vartheta_i$ for $+$ $d$-simplex $t$ is defined by first assigning even $(d-1)$-simplices in ascending order, then odd simplices in ascending order again:
\begin{equation}
    0\to 2\to 4\to\dots \to 1\to 3\to 5\to\dots
\end{equation}
\item For $-$ $d$-simplices, the order is defined in opposite way:
\begin{equation}
    \dots\to 5\to 3\to 1 \to \dots \to 4\to 2\to 0.
\end{equation}
\end{itemize}
For example, for $d=2$, $u(012)=\vartheta_{12}^{\alpha(12)}\vartheta_{01}^{\alpha(01)}\vartheta_{02}^{\alpha(02)}$ when $(012)$ is a $+$ triangle, 
and $u(012)=\vartheta_{02}^{\alpha(02)}\vartheta_{01}^{\alpha(01)}\vartheta_{12}^{\alpha(12)}$ for a $-$ triangle. 
Then, We choose the assignment of $\theta$ and $\overline{\theta}$ on each $e$ such that, if $t$ is a $+$ (resp.~$-$) simplex, $u(t)$ includes $\overline{\theta}_e$ when $e$ is labeled by an odd (resp.~even) number, see Fig.~\ref{fig:Grassmann}.

Based on the above definition of $u(t)$, the quadratic property of $u(t)$ is given by
\begin{equation}
    \sigma(\alpha)\sigma(\alpha')=\sigma(\alpha+\alpha')(-1)^{\int\alpha\cup_{d-2}\alpha'},
    \label{eq:ClosedQuad}
\end{equation}
for closed $\alpha, \alpha'$. To see this, we just have to bring the product of two Grassmann integrals
\begin{equation}
    \sigma(\alpha)\sigma(\alpha')=\int\prod_{e|\alpha(e)=1}d\theta_e d\overline{\theta}_e\prod_{e|\alpha'(e)=1}d\theta_e d\overline{\theta}_e \prod_t u(t)[\alpha]\prod_t u(t)[\alpha']
\end{equation}
into the form of $\sigma(\alpha+\alpha')$ by permuting Grassmann variables, and count the net fermionic sign.
First of all, each path integral measure on $e$ picks up a
sign $(-1)^{\alpha(e)\alpha'(e)}$ by permuting $d\overline{\theta}_e^{\alpha(e)}$ and $d\theta_e^{\alpha'(e)}$.
For integrands, $u(t)$ on different $d$-simplices commute with each other for closed $\alpha$, so nontrivial signs occur only by reordering $u(t)[\alpha]u(t)[\alpha']$ to $u(t)[\alpha+\alpha']$ on a single $d$-simplex. The sign on $t$ is explicitly written as
\begin{equation}
    (-1)^{\sum_{e,e'\in t}^{e>e'}\alpha(e)\alpha'(e')},
\end{equation}
where the order $e>e'$ is determined by $u(t)$. Hence, the net fermionic sign is given by
\begin{equation}
    \sigma(\alpha)\sigma(\alpha')=\sigma(\alpha+\alpha')\prod_t(-1)^{\epsilon[t, \alpha, \alpha']},
    \label{eq:orientedquad}
\end{equation}
with
\begin{equation}
   \epsilon[t, \alpha, \alpha']=\sum_{e,e'\in t, e>e'}\alpha(e)\alpha'(e')+\sum_{e\in t, e>0}\alpha(e)\alpha'(e),
   \label{eq:redis}
\end{equation}
where $e>0$ if $u[t]$ includes a $\overline{\theta}_e$ variable. Then, the sign $\epsilon[t, \alpha, \alpha']$ has a neat expression in terms of the higher cup product.
For later convenience, we compute $\epsilon[t, \alpha, \alpha']$ including the case that $\alpha, \alpha'$ are not closed. 

At a $+$ simplex, after some efforts we can rewrite $\epsilon[t, \alpha, \alpha']$ as
\begin{equation}
\begin{split}
    \epsilon[t, \alpha, \alpha']&=\sum_{i}\alpha_{2i+1}\cdot \delta\alpha'(t)+\sum_{i<j}\alpha_{2i+1}\alpha'_{2j+1}+\sum_{i>j}\alpha_{2i}\alpha'_{2j}\\
    &=\alpha\cup_{d-2}\alpha'+\alpha\cup_{d-1}\delta\alpha'.
    \end{split}
    \label{eq:pepsilon}
\end{equation}

At a $-$ simplex, similarly we have
\begin{equation}
\begin{split}
    \epsilon[t, \alpha, \alpha']&=\sum_{i}\alpha_{2i}\cdot \delta\alpha'(t)+\sum_{i<j}\alpha_{2i+1}\alpha'_{2j+1}+\sum_{i>j}\alpha_{2i}\alpha'_{2j}\\
    &=\delta\alpha(t)\delta\alpha'(t)+\alpha\cup_{d-2}\alpha'+\alpha\cup_{d-1}\delta\alpha'.
    \end{split}
    \label{eq:mepsilon}
\end{equation}
We can see the quadratic property~\eqref{eq:ClosedQuad} when $\alpha, \alpha'$ are closed. 

The change of $\sigma(\alpha)$ under the gauge transformation
$\alpha\to \alpha+\delta \gamma$  or under the change of the triangulation is controlled by the formula \begin{equation}
\sigma(\tilde M,\tilde \alpha) 
= (-1)^{\int_K (\Sq^2 (\alpha) + w_2 \cup \alpha)} \sigma(M,\alpha), 
\end{equation}
where $\tilde M$ is the same manifold $M$ with a different triangulation,
$\tilde\alpha$ is a cocycle such that $[\alpha]=[\tilde\alpha] $ in cohomology,
and $K=M\times [0,1]$ such that the two boundaries are given by $M$ and $\tilde M$,
and finally $\alpha$ is extended to $K$ so that it restricts to $\alpha$ and $\tilde \alpha$ on the boundaries.
The derivation was given in \cite{Gaiotto:2015zta}.

We note that due to the Wu relation~\cite{ManifoldAtlasWu}, we have \begin{equation}
(-1)^{\int_K (\Sq^2(\alpha) + w_2\cup\alpha)} = +1,
\end{equation}  when $K$ is an oriented closed manifold and $\alpha$ is a cocycle.
This means that $\int_K (\Sq^2 (\alpha) + w_2\cup\alpha)$ represents a trivial phase in $d+1$ dimensions,
and therefore there should be a trivial boundary in $d$ dimensions.
We can think of the Gu-Wen Grassmann integral $\sigma(M,\alpha)$ as providing an explicit formula for such a trivial boundary. 

\subsection{Gu-Wen spin $G$-SPT phase}
The Gu-Wen spin invertible theories form a subgroup of $\Hom(\Omega^\text{spin}_{d}(BG),U(1))$ and is specified by a pair $(m_{d-1},x_{d})\in Z^{d-1}(BG,\bZ_2)\times C^{d}(BG,U(1))$ satisfying $\Sq^2 (m_{d-1}) = \delta x_{d}$, where $\Sq^2(m) := m \cup_{d-2} m$.
For a given $g:M\to BG$ where $M$ is a spin $d$-manifold, the action of the invertible theory is given by \cite{Gu:2012ib,Gaiotto:2015zta}\footnote{%
For a more mathematical treatment, see papers by Brumfiel and Morgan \cite{BFquadratic}.
} \begin{equation}
\sigma(g^*m_{d-1})  \exp(\pi i \int_M(\eta\cup g^*m_{d-1} + g^*x_d))
\end{equation}
where $\sigma(g^*m_{d-1})=\pm1$ is the Grassmann integral of Gu-Wen \cite{Gu:2012ib} as formulated by Gaiotto and Kapustin \cite{Gaiotto:2015zta},
and $\delta\eta=w_2$ specifies the chosen spin structure.

\section{Grassmann integral for pin case}
\label{sec:pinGuWen}
Now let us construct the Grassmann integral $\sigma(M, \alpha)$ on a $d$-manifold $M$ which might be unoriented. 
We construct an unoriented manifold by picking locally oriented patches, and then gluing them along codimension one loci by transition functions. The locus where the transition functions are orientation reversing, constitutes a representative of the dual of first Stiefel-Whitney class $w_1$. We will sometimes call the locus an orientation reversing wall.
Again, we endow $M$ with a barycentric subdivision for the triangulation of $M$. We then assign a local ordering to vertices of the barycentric subdivision, such that a vertex on the barycenter of $i$ vertices is labeled as $i$.

\begin{figure}[htb]
\centering
\includegraphics{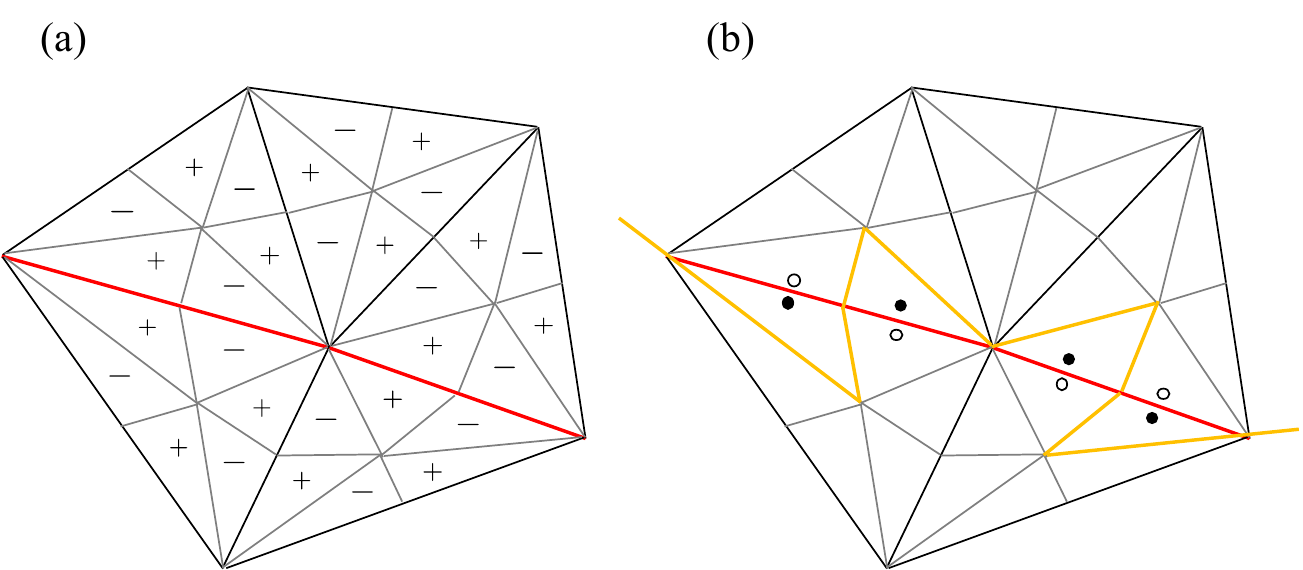}
\caption{(a): The signs of $d$-simplices near the orientation reversing wall, which is represented as a red line. (b): Assignment of Grassmann variables on the wall specifies a deformation of the wall that intersects the wall transversally at $(d-2)$-simplices.
}
\label{fig:wall}
\end{figure}

For the oriented case, we have placed a pair of Grassmann variables $\theta_e, \overline{\theta}_e$ on each $(d-1)$-simplex $e$, whose assignment is determined by the sign of $d$-simplices ($+,-$) sharing $e$. We remark that the assigning rule fails, when $e$ lies on the wall where we glue patches of $M$ by the orientation reversing map. In this case, we would have to assign Grassmann variables of the same color on both sides of $e$ (i.e., both are black ($\theta$) or white ($\overline{\theta}$)), since the two $d$-simplices sharing $e$ have the identical sign when $e$ is on the orientation reversing wall, see Fig.~\ref{fig:wall} (a). Hence, we need to slightly modify the construction of the Grassmann integral on the orientation reversing wall. To do this, instead of specifying a canonical rule to assign Grassmann variables on the wall,
we just place a pair $\theta_e$, $\overline{\theta}_{e}$ on the wall in an arbitrary fashion. 
Then, we define the Grassmann integral as
\begin{equation}
    \sigma(M, \alpha)=\int\prod_{e|\alpha(e)=1}d\theta_e d\overline{\theta}_e \prod_t u(t)\prod_{e|\mathrm{wall}}(\pm i)^{\alpha(e)},
    \label{sigmadefpin}
\end{equation}
where the $\prod_{e|\mathrm{wall}}(\pm i)^{\alpha(e)}$ term assigns weight $+i^{\alpha(e)}$ (resp.~$-i^{\alpha(e)}$) on each $(d-1)$-simplex $e$ on the orientation reversing wall, when $e$ is shared with $+$ (resp.~$-$) $d$-simplices. There is no ambiguity in such definition, since both $d$-simplices on the side of $e$ have the same sign. This factor makes the Grassmann integral a $\bZ_4$ valued quadratic function. The quadratic property is expressed as
\begin{equation}
    \sigma(\alpha)\sigma(\alpha')=\sigma(\alpha+\alpha')(-1)^{\int\alpha\cup_{d-2}\alpha'}.
    \label{eq:unorientedquad}
\end{equation}
Basically, the quadratic property is derived in the similar fashion to the oriented spin case. In this case, the net sign consists of three parts;
\begin{itemize}
    \item the fermionic sign that occurs when reordering $u(t)[\alpha]u(t)[\alpha']$ to $u(t)[\alpha+\alpha']$ on a single $d$-simplex. The sign on $t$ is expressed as
    \begin{equation}
    (-1)^{\sum_{e,e'\in t}^{e>e'}\alpha(e)\alpha'(e')}.
\end{equation}
\item the fermionic sign by permuting the path integral measure, $(-1)^{\alpha(e)\alpha'(e)}$ on each $(d-1)$-simplex.
\item the sign that comes from $i^{\alpha(e)}$ factor on the wall, which is given by comparing $ i^{\alpha(e)} i^{\alpha'(e)}$ with $i^{\alpha(e)+\alpha'(e)}$, with the sum of $\alpha$ taken mod 2. This part counts $(-1)^{\alpha(e)\alpha'(e)}$ on the orientation reversing wall.
\end{itemize}
Analogously to what we did to the second term in~\eqref{eq:redis} for
the oriented case, we try to re-distribute the fermionic sign from the measure $\prod_{e}(-1)^{\alpha(e)\alpha'(e)}$ to $d$-simplices, by assigning $(-1)^{\alpha(e)\alpha'(e)}$ to a $+$ simplex (resp.~$-$ simplex) $t$ sharing $e$, when $e$ is labeled by an odd (resp.~even) number. 
However, such a distribution fails when $e$ is on the orientation reversing wall, due to the mismatch of the sign of two $d$-simplices on the side of $e$. Such a distribution counts no sign on the orientation reversing wall. 
But, this lack of the sign on the wall is complemented by 
the factor $(-1)^{\alpha(e)\alpha'(e)}$
from the contribution of the
$i^{\alpha(e)}$ term, making the re-distribution possible after all. Hence, we can express the net sign in exactly the
same fashion as the oriented case~\eqref{eq:orientedquad}, which proves~\eqref{eq:unorientedquad}.

\subsection{Effect of re-triangulation}
\label{sec:retri}
\begin{figure}[htb]
\centering
\includegraphics{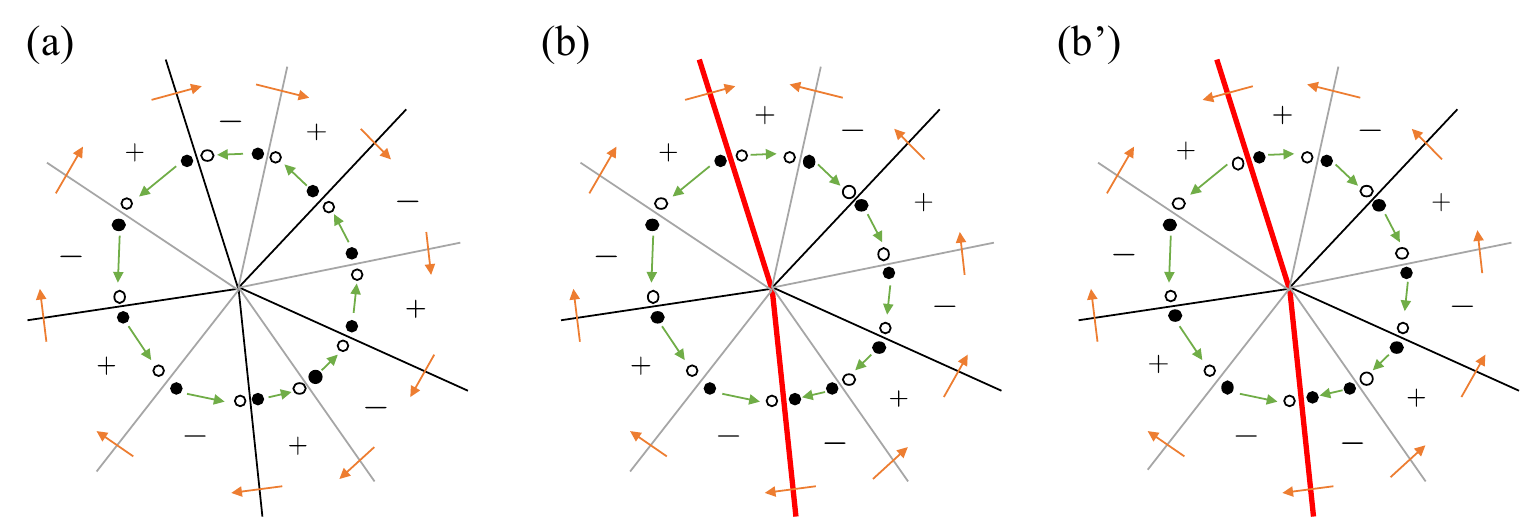}
\caption{When $\lambda(v)=1$ on a single $(d-2)$ simplex $v$, Grassmann variables on $(d-1)$-simplices surrounding $v$ are counted in the integral. In the expression of the integral, we encounter  $\pm d\vartheta_{2i}d\vartheta_{2i+1}$ measure factors from $(d-1)$-simplices, and $\pm \vartheta_{2i+1}\vartheta_{2i+2}$ integrand factors from $d$-simplices. The sign $\pm$ from the measure (resp.~integrand) is expressed by the orange (resp.~green) arrow. For instance, the arrow is directed from $\vartheta_{2i}$ to $\vartheta_{2i+1}$ if we have a $+$ sign on the measure, otherwise directed in the opposite direction.
(a): If $v$ is away from the orientation reversing wall, we can see that all the signs from the measure share the same sign. We can also check that signs from the integrand have the same sign. In such a situation, we have $\sigma(\delta\lambda)=-1$. (b): If $v$ is placed on the orientation reversing wall (red thick line), we have to flip the direction of all arrows on one side of the wall. The total number of flipped arrows is odd; odd number of orange arrows and even number of green arrows. Thus, the value of the integral in (b) has the opposite sign from that of (a). Hence, we have $\sigma(\delta\lambda)=+1$, when the two Grassmann variables attached on one side of the wall have different colors. (b'): On the other hand, we have $\sigma(\delta\lambda)=-1$, when the two Grassmann variables attached on one side of the wall have the
same color.}

\label{fig:stiefel}
\end{figure}

Next, we move on
to discuss the effect of re-triangulation. Suppose we have two configurations of $\alpha$, orientation reversing walls and triangulations on $M\times\{0\}$ and $M\times\{1\}$. Then, we will see that
\begin{equation}
    \sigma(M\times\{0\})=(-1)^{\int_{K}(\Sq^2(\alpha)+(w_2+w_1^2)\cup\alpha)}\sigma(M\times\{1\}),
    \label{eq:pinretriangulate}
\end{equation}
where $K=M\times[0,1]$, and $\alpha$ on $M\times\{0\}$, $M\times\{1\}$ is extended to $K$. To see this, we first observe the quadratic property of $\tilde{\sigma}(\alpha):=\sigma(M\times\{0\})\sigma(M\times\{1\})^{-1}$,
\begin{equation}
    \tilde{\sigma}(\alpha)\tilde{\sigma}(\alpha')=\tilde{\sigma}(\alpha+\alpha')(-1)^{\int_{\partial K}\alpha\cup_{d-2}\alpha'}.
    \label{eq:MMquad}
\end{equation}
Since~\eqref{eq:MMquad} is satisfied for $\tilde{\sigma}'(\alpha)=(-1)^{\int_K \Sq^2(\alpha)}$,
 we can express $\tilde{\sigma}(\alpha)$ as $(-1)^{\int_K \Sq^2(\alpha)}$ up to linear term,
 \begin{equation}
     \tilde{\sigma}(\alpha)=(-1)^{\int_K \Sq^2(\alpha)}(-1)^{\sum_{e\in K}\chi(e)\alpha(e)}.
 \end{equation}
 The linear term is fixed by computing $\tilde{\sigma}(\alpha)$ in the simplest case; $\alpha=\delta\lambda$ on $\partial K$, and $\lambda(v)=1$ on a single $(d-2)$-simplex of $\partial K$, otherwise 0. Once we take 
a barycentric subdivision, when $\lambda$ is nonzero away from the orientation reversing wall, one can see that $\tilde{\sigma}(\delta\lambda)=-1$,
 by imitating the logic of Sec.~4.1.~of Gaiotto-Kapustin~\cite{Gaiotto:2015zta}. See also Fig.~\ref{fig:stiefel} (a).
 In the case that $\lambda$ is nonzero on the orientation reversing wall, the value of $\tilde{\sigma}(\delta\lambda)$ depends on the way of assigning Grassmann variables to $(d-1)$-simplices on the wall such that $\delta\lambda=1$. For simplicity, we examine the case that $\delta\lambda$ is nonzero on two $(d-1)$-simplices on the wall.
 (In general, there are even number of such $(d-1)$-simplices. It is not hard to generalize for these situations.)
 Then, we have two Grassmann variables attached on each side of the orientation reversing wall. When the two Grassmann variables on one side of the wall share the same color (i.e., both are black ($\theta$) or white ($\overline{\theta}$)), we can show that $\tilde{\sigma}(\delta\lambda)=-1$ (see Fig.~\ref{fig:stiefel} (b')).
 
 On the other hand, if the Grassmann variables on one side have 
different colors (i.e., one $\theta$ and one $\overline{\theta}$), we have $\tilde{\sigma}(\delta\lambda)=+1$ (see Fig.~\ref{fig:stiefel} (b)).
 (In these computations, the $\prod (\pm i)^{\delta\lambda(e)}$ term spits no sign, $(+i)\cdot(-i)=1$.)
 
 Now let us determine the linear term. First, let us recall that the set of all $(d-2)$-simplices of the barycentric subdivision 
gives the representative of the dual of $w_2$. 
Thus, we can express $\tilde{\sigma}(\alpha)$ as
  \begin{equation}
     \tilde{\sigma}(\alpha)=(-1)^{\int_K (\Sq^2(\alpha)+w_2\cup\alpha)}(-1)^{\sum_{e\in K}\chi'(e)\alpha(e)}.
 \end{equation}
Here, $(-1)^{\sum_{e\in K}\chi'(e)\alpha(e)}=1$ if $\lambda$ is nonzero away from the orientation reversing wall. When $\lambda$ is nonzero on the wall, $(-1)^{\sum_{e\in K}\chi'(e)\alpha(e)}=1$ (resp.~$-1$) if the two Grassmann variables on one side of the wall have the same (resp.~different) color. We can express such a linear term as $(-1)^{\int_K w_1^2\cup\alpha}$. To see this, first we observe that the choice of the assignment of Grassmann variables on the wall corresponds to choosing the slight deformation of the wall, such that the deformation intersects transversally with the wall at $(d-2)$-simplices. Concretely, we deform the wall on each $(d-1)$-simplices of the wall to the side where $\theta$ (black dot) is contained, see Fig.~\ref{fig:wall} (b). Now we can see that $(-1)^{\sum_{e\in K}\chi'(e)\alpha(e)}=-1$ when $\lambda=1$ at the intersection of these two walls, otherwise 1. Here, both walls before and after deformation give a representative of the dual of $w_1$, and thus the intersection of two walls gives a representative of the dual of $w_1^2$. Hence, we have $(-1)^{\sum_{e\in K}\chi'(e)\alpha(e)}=(-1)^{\int_K w_1^2\cup\alpha}$, proving~\eqref{eq:pinretriangulate}.

 \subsection{Gu-Wen pin SPT phase}
In this subsection, we discuss the fermionic SPT phases on an unoriented spacetime.
To do this, let us begin with recalling the construction of bosonic SPT phases on unoriented manifolds, following~\cite{Bhardwaj2017}. Here, we limit ourselves to the case that the structure group is decomposed as $G_0\times O(d)$. Then, the $G_0$ connection $g_0: M\to BG_0$ together with $w_1$ defines a connection of $G_0\times \bZ_2^R$, $g: M\to B(G_0\times \bZ_2^R)$, where $\bZ_2^R$ is the $\bZ_2$ subgroup of $O(d)$ generated by the orientation reversing element.
From now, we will simply write $G:=G_0\times\bZ_2^R$.
We denote $\rho$ as a $G$-action on $U(1)$, such that $\rho_g(a)=a^{r(g)}$ for $g\in G, a\in U(1)$, where $r(g)=-1$ when $g\in G$ reverses the orientation, otherwise $+1$.

Then, a well-understood class of $d$-dimensional bosonic SPT phases is classified
by the $\rho$-twisted cohomology group $H^d(BG, U(1)_{\rho})$~\cite{ChenGuLiuWen2011}. 
For a given $\omega\in Z^d(BG, U(1)_{\rho})$, the action of the SPT phase on an unoriented $d$-manifold $M$ is given by a certain product of weights $g^*\omega$ on each $d$-simplex of $M$, which is constructed as follows.

First, let us consider the case that the $d$-simplex $t$ is away from the orientation reversing wall.
In this case, we simply define the weight as $g^*\omega^{s(t)}$, where $s(t)=+1$ if $t$ is a $+$ simplex, and $s(t)=-1$ if $t$ is a $-$ simplex, which is identical to the definition of the oriented case. 
However, when the $d$-simplex $t$ traverses the orientation reversing wall, the definition of the weight described above should be modified, since the choice of the sign $s(t)$ has an ambiguity.
To resolve such ambiguity, we first assign $+1$ to every vertex of $t$ on one side of the wall, and assign $-1$ on the other side. Then, we define $s(t)$ as the sign given by comparing the ordering on $t$ and the orientation of $M$ on the side of vertices labeled by $+1$.
Let us denote $\epsilon$ as the number $\pm1$ assigned on the vertex of the smallest ordering in $t$. 
Then, we define the weight on $t$ as $g^*\omega^{\epsilon\cdot s(t)}$. 

We note that this definition is independent of the choice of assigning $\pm 1$ to one side of the wall, since flipping the sign of $\pm 1$ on vertices changes the sign of $\epsilon$ and $s(t)$ simultaneously, which leaves $g^*\omega^{\epsilon\cdot s(t)}$ invariant. 
Then, let us write the action as the product of weights for all $d$-simplices in $M$. We simply denote the action as $\int_M g^*\omega$. One can see that such defined action is invariant under re-triangulation~\cite{Bhardwaj2017}. If we take a general cochain $x\in C^{d}(BG, U(1)_{\rho})$ which is not necessarily a cocycle, we can see that $\int_M g^*x$ is no longer invariant under re-triangulation, whose variation is controlled by $\int g^*(\delta_{\rho}x)$.

Now, we are ready to consider the fermionic case.
 The Gu-Wen SPT phase based on pin$_{-}$ structure is specified by a pair $(m_{d-1},x_{d})\in Z^{d-1}(BG,\bZ_2)\times C^{d}(BG,U(1)_{\rho})$ satisfying $\Sq^2 m_{d-1} = \delta_{\rho} x_{d}$.
 For a given $g:M\to BG$ where $M$ is a pin$_{-}$ $d$-manifold, the action of the invertible theory is given by 
\begin{equation}
\sigma(g^*m_{d-1})  \exp(\pi i \int_M(\eta\cup g^*m_{d-1} + g^*x_d)),
\label{eq:guwenpin}
\end{equation}
where $\delta\eta=w_2+w_1^2$ specifies the chosen pin$_{-}$ structure.

On the other hand, the pin$_+$ Gu-Wen SPT phase is given by $(m_{d-1}, x_d)$ such that  $\Sq^2 m_{d-1}+\rho_1^2\cup m_{d-1} = \delta_{\rho} x_d$. 
Here, we define $\rho_1\in Z^1(BG,\bZ_2)$ such that $w_1=g^*\rho_1$, as a map that sends $\bZ_2^R$ odd element of $G$ to 1, otherwise 0.  
Then, the action of the invertible theory is given in the form of~\eqref{eq:guwenpin}, where $\delta\eta=w_2$ specifies the chosen pin$_{+}$ structure.

\section{Arf-Brown-Kervaire invariant in (1+1)d}
\label{sec:ABK}
In this section, we construct the 2d pin$_{-}$ invertible TQFT~\cite{Debray2018} for the
Arf-Brown-Kervaire (ABK) invariant via the Grassmann integral on lattice,  whose state sum definition was initially given in~\cite{Turzillo2018}. 
In condensed matter literature, this invertible theory describes
(1+1)d topological superconductors
in class BDI~\cite{FidkowskiKitaev2011}. 
Here, we construct the $\bZ_8$-valued ABK invariant by coupling the 2d state sum shadow TQFT with the Grassmann integral, which was performed for the $\bZ_2$-valued Arf invariant of the spin case in~\cite{Gaiotto:2015zta}. 

The weight for the state sum is assigned in the same manner as the case of the Arf invariant of the spin case~\cite{Gaiotto:2015zta}, described as follows. 
For a given configuration
$\alpha\in C^1(M, \bZ_2)$, we assign weight $1/2$ to each 1-simplex $e$, and also assign weight $2$ to each 2-simplex $f$ when $\delta\alpha=0$ at $f$, otherwise 0. Let us denote the product of the whole weight as $\tilde{Z}[\alpha]$.
Then, we can see that the partition function is given by the ABK invariant up to Euler term, 
\begin{equation}
\begin{split}
    Z[M,\eta]&=\sum_{\alpha\in Z^1(M,\bZ_2)}\sigma(M,\alpha)(-1)^{\int_M \eta\cup\alpha}\tilde{Z}[\alpha] \\
    &=2^{|F|-|E|}\cdot\sum_{\alpha\in Z^1(M,\bZ_2)}\sigma(M,\alpha)(-1)^{\int_M \eta\cup\alpha}\\
    &=2^{\chi(M)-1}\cdot\sum_{[\alpha]\in H^1(M,\bZ_2)}\sigma(M,\alpha)(-1)^{\int_M \eta\cup\alpha}\\
    &=\sqrt{2}^{\chi(M)}\mathrm{ABK}[M,\eta],
    \end{split}
\end{equation}
where $|F|$, $|E|$ denotes the number of 2-simplices, 1-simplices in $M$, respectively. $\chi(M)$ denotes the Euler number of $M$, and ABK$[M,\eta]$ is the
ABK invariant,
\begin{align}
    \mathrm{ABK}[M,\eta]=\frac{1}{\sqrt{|H^1(M,\bZ_2)|}}\sum_{[\alpha]\in H^1(M,\bZ_2)}i^{Q_{\eta}[\alpha]}.
\end{align}
Here, $i^{Q_{\eta}[\alpha]}=\sigma(M,\alpha)(-1)^{\int_M \eta\cup\alpha}$ is a $\bZ_4$-valued quadratic function that satisfies
\begin{equation}
    Q_{\eta}[\alpha]+Q_{\eta}[\alpha']=Q_{\eta}[\alpha+\alpha']+2\int_M\alpha\cup\alpha'.
    \label{eq:refine}
\end{equation}

The ABK invariant determines the pin$_-$ bordism class of 2d manifolds $\Omega_2^{\mathrm{pin}_-}(\mathrm{pt})=\bZ_8$, which is generated by $\mathbb{RP}^2$~\cite{KirbyTaylor}. 
To see this, let $\alpha$ be
a nontrivial 1-cocyle that generates $H^1(\mathbb{RP}^2, \bZ_2)=\bZ_2$. Then, using the quadratic property for $\alpha=\alpha'$ in~\eqref{eq:refine}, one can see that $Q_{\eta}[\alpha]$ takes value in $\pm 1$, since $Q_{\eta}[0]=0$ and $\int_M \alpha\cup\alpha'=1$.  
$Q_{\eta}[\alpha]=\pm 1$ corresponds to two possible choices of pin$_{-}$ structure on $\mathbb{RP}^2$. Then, the ABK invariant is computed as an 8th root of unity,
\begin{equation}
\mathrm{ABK}[M, \eta]=\frac{1\pm i}{\sqrt{2}}=e^{\pm 2\pi i/8}.
\end{equation}

\section{Gapped boundary of Gu-Wen pin SPT phase}
\label{sec:boundary}
In this section, we demonstrate that Gu-Wen pin $G$-SPT phases admit a
gapped boundary, by writing down the explicit $d$ dimensional action on the boundary of $(d+1)$ dimensional Gu-Wen pin $G$-SPT phase specified by the Gu-Wen data $(n_d,y_{d+1})$.
To construct the gapped boundary, we prepare a symmetry extension by a (0-form) symmetry $\tilde{K}$~\cite{Wang:2017loc},
\begin{equation}
0\to \tilde{K}\to \tilde{H} \stackrel{\tilde{p}}{\to} G \to 0, 
\end{equation}
such that $n_d$ trivializes as an element of $H^d(B\tilde{H}, \bZ_2)$; 
$[\tilde{p}^*n_d]=0\in H^d(B\tilde{H}, \bZ_2)$. 
When $G$ is finite, such an extension can be prepared by generalizing the argument of \cite{Tachikawa:2017gyf}.

We now take $\tilde{m}_{d-1}\in C^{d-1}(B\tilde{H}, \bZ_2)$ such that $\tilde{p}^*n_d=\delta \tilde{m}_{d-1}$.
In pin$_-$ case, we see that $z_{d+1}=\tilde{p}^*y_{d+1}-\Sq^2 (\tilde{m}_{d-1})$ is a ($\rho$-twisted) cocycle, where $\Sq^2 (\tilde{m}_{d-1})=\tilde{m}_{d-1}\cup_{d-3} \tilde{m}_{d-1} + \delta \tilde{m}_{d-1} \cup_{d-2} \tilde{m}_{d-1}$.
Therefore, the bulk Gu-Wen data pull back to
$(\delta \tilde{m}_{d-1}, \Sq^2 (\tilde{m}_{d-1})+ z_{d+1})$.
In pin$_+$ case, we instead define the $\rho$-twisted cocycle $z_{d+1}=\tilde{p}^*y_{d+1}-\Sq^2 (\tilde{m}_{d-1})-(\tilde{p}^*\rho_1)^2\cup \tilde{m}_{d+1}$.
Then, one can see that the Gu-Wen data pull back to $(\delta \tilde{m}_{d-1}, \Sq^2 (\tilde{m}_{d-1})+(\tilde{p}^*\rho_1)^2 \cup \tilde{m}_{d-1}+ z_{d+1})$.

Without loss of generality we can assume that $z_{d+1}=\delta_{\rho} x_d$ for some $x_d\in C^{d}(B\tilde{H},U(1)_{\rho})$,
by a further extension of the symmetry 
\begin{equation}
0\to K \to H \stackrel{p}{\to} \tilde{H} \to 0.
\end{equation}
Again, such an extension for twisted cocycle can be prepared by generalizing the argument of \cite{Tachikawa:2017gyf}. 
We set $m_{d-1}=p^*\tilde{m}_{d-1}$.
We now expect that the action on the boundary is given by the $K$-gauge theory,
\begin{equation}
Z_\text{boundary gauge}\propto \sum_{p(h)=g} \sigma(h^*m_{d-1}) \exp(\pi i \int_M (\eta\cup h^*m_{d-1} + h^*x_d)),
\label{gwb}
\end{equation}
with $h: M\to BH$.
But to make sense of this expression we have to extend the definition of the Gu-Wen Grassmann integral $\sigma(\alpha_{d-1})$ to the case when $\alpha_{d-1}\in C^{d-1}(M, \bZ_2)$ is not necessarily closed. This generalization was performed for the spin case in~\cite{KOT2019}.
By slightly generalizing the analysis in~\cite{KOT2019} to the pin case, we will see that the extended Gu-Wen integral nicely couples to the bulk in a gauge invariant fashion.
\subsection{Bulk-boundary Gu-Wen Grassmann integral for the pin case}
When we naively use the above definition \eqref{sigmadef} when  $\alpha$ is not closed: $\delta \alpha=\beta$, 
the resulting expression is problematic since $u(t)$ can become Grassmann-odd.
Following~\cite{KOT2019}, we avoid this conundrum by coupling with the Gu-Wen integral $\sigma(N,\beta)$ in $(d+1)$ dimensional bulk $N$ such that $\partial N=M$, making all components in the path integral Grassmann-even. 

Now let us write down the boundary Gu-Wen integral coupled with bulk;
we denote the entire integral by $\sigma(\alpha;\beta)$. 
We assign Grassmann variables $\theta_e, \overline{\theta}_e$ on each $(d-1)$-simplex $e$ of $M$, and $\theta_f, \overline{\theta}_f$ on each $d$-simplex $f$ of $N\setminus M$.
We define the Gu-Wen integral as
\begin{equation}
    \sigma(\alpha;\beta)=\int\prod_{f|\beta(f)=1}d\theta_f d\overline{\theta}_f \int\prod_{e|\alpha(e)=1}d\theta_e d\overline{\theta}_e \prod_t u(t)\prod_{f|\mathrm{wall}}(\pm i)^{\beta(f)}\prod_{e|\mathrm{wall}}(\pm i)^{\alpha(e)},
    \label{eq:GWboundary}
\end{equation}
where we assume that the orientation reversing wall in $N$ intersects $M$ transversally at $(d-1)$-simplices, which are regarded as making up the wall in $M$.
$u(t)$ is a monomial of Grassmann variables defined on a $(d+1)$-simplex of $N$.  
$u(t)[\beta]$ is defined in the same fashion as in the case without boundary if $t$ is away from the boundary, but modified when $t$ shares a $d$-simplex with the boundary. 
For simplicity, we assign an ordering on vertices of such $t=(01\dots d+1)$, so that the $d$-simplex shared with $M$ becomes $f_0=(12\dots d+1)$; the vertex $0$ is contained in $N\setminus M$. 
For instance, we can take a barycentric subdivision on $N$, and assign $0$ to vertices associated with $(d+1)$-simplices. 
We further define the sign of $d$-simplices on $M$, such that $f_0$ and $t$ have the same sign.

Then, $u(t)$ neighboring with $M$ is defined by replacing the position of $\vartheta_{f_0}$ in $u(t)[\beta]$ with the boundary action on $f_0$, $u(f_0)[\alpha]=\prod_{e\in f_0}\vartheta_e^{\alpha(e)}$. 
We then have: On a $+$ simplex,
\begin{equation}
    u(t)=u(f_0)[\alpha]\cdot\prod_{f\in\partial t, f\neq f_0}\vartheta_f^{\beta(f)}.
\end{equation}
On a $-$ simplex,
\begin{equation}
    u(t)=\prod_{f\in\partial t, f\neq f_0}\vartheta_f^{\beta(f)}\cdot u(f_0)[\alpha].
\end{equation}
One can check that $u(t)$ defined above becomes Grassmann-even. Then, using the exactly same logic as Sec.~4.3.~of~\cite{KOT2019}, one can obtain the quadratic property of $\sigma(\alpha; \beta)$ as
\begin{equation}
    \sigma(\alpha+\alpha'; \beta+\beta')=\sigma(\alpha;\beta)\sigma(\alpha';\beta')(-1)^{\int_M (\alpha\cup_{d-2}\alpha'+\alpha\cup_{d-1}\delta\alpha')+\int_N \beta\cup_{d-1}\beta'}.
    \label{eq:Quadboundary}
\end{equation}

\subsection{Effect of re-triangulation}
\begin{figure}[htb]
\centering
\includegraphics{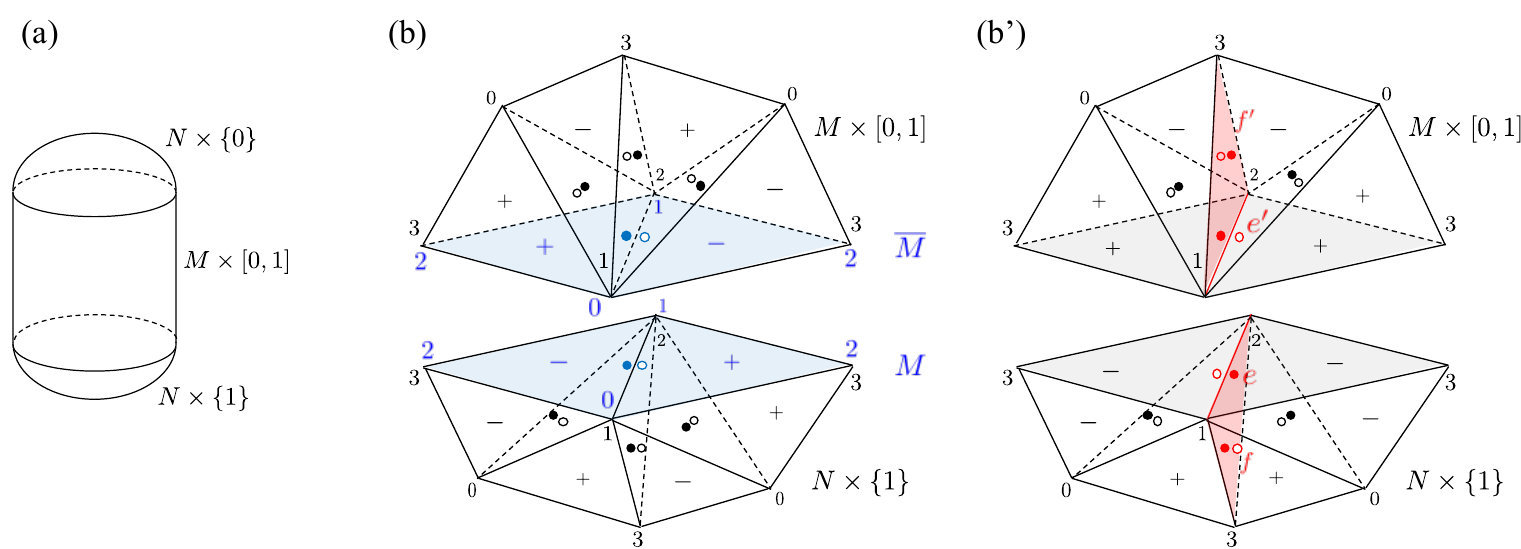}
\caption{(a): An example of $K$ such that $\p K=(N\times\{0\})\sqcup (M\times[0,1])\sqcup (N\times\{1\})$. (b): Away from the orientation reversing wall, the colors of Grassmann variables on $M$ match those of $\overline{M}$. (b'): On the orientation reversing wall (red plane), the Grassmann variable is assigned such that (1) the color of the Grassmann variable of $e'$ on one side of the wall is the same as that of $f$, and (2) the color of the Grassmann variable of $e$ on one side of the wall is different from that of $e'$.}
\label{fig:attach}
\end{figure}

We can determine the effect of re-triangulation 
on $\sigma(\alpha;\beta)$ from the quadratic property. To compare the value of the Gu-Wen integral on $N$ with different triangulations, we consider $K=N\times[0, 1]$, with the Gu-Wen integral on $\partial K= (N\times\{0\})\sqcup (M\times[0,1])\sqcup (N\times\{1\})$, see Fig.~\ref{fig:attach} (a).
Suppose we have two triangulations and configurations of $(\alpha, \beta)$ we want to compare, on $N\times\{0\}$ and $N\times\{1\}$, respectively. 
Roughly speaking, we will compute the effect of re-triangulations by showing that
\begin{equation}
    \sigma(N\times\{0\})\sigma(M\times[0,1])\sigma^{-1}(N\times\{1\})
    =(-1)^{\int_K(\Sq^2(\beta)+(w_2+w_1^2)\cup\beta)},
    \label{eq:retriangle}
\end{equation}
and
\begin{equation}
\sigma(M\times [0,1])=(-1)^{\int_{M\times[0,1]}
(\Sq^2(\alpha)+(w_2+w_1^2)\cup\alpha)}.
\label{eq:M01}
\end{equation}
To demonstrate these relations, we need to 
modify slightly the definition of $\sigma(N\times\{0\})$ and
$\sigma(N\times\{1\})$ on the boundary $M\sqcup \overline{M}$. First, for $(d-1)$-simplices $e$ in $M\sqcup \overline{M}$, we change the role of $\theta_e$ and $\overline{\theta}_e$ in integrands, when $e$ is not contained in the orientation reversing wall. By this redefinition, the color of Grassmann variables away from the wall in $\partial(N\times\{0\})\sqcup\partial(N\times\{1\})$ match with that of $\partial (M\times[0,1])$, see Fig.~\ref{fig:attach} (b). 
Such a redefinition changes $\sigma(N)$ only by a linear and gauge invariant counterterm,
\begin{equation}
\sigma(N)\to \sigma(N)\cdot (-1)^{\sum_{f_{+}\in M}\beta(f_+)},
\end{equation}
where $f_{+}$ denotes $+$ simplices on $M$.

In addition, when $e$ is placed on the wall, we have to choose the assignment of Grassmann variables deliberately.
Concretely, let $e$ be a $(d-1)$ simplex of $M\times\{0\}$ on the side of $N\times\{0\}$. Then, we denote $e'$ as a $(d-1)$ simplex of $M\times\{0\}$ on the side of $M\times[0,1]$, which matches with $e$ by gluing $N\times\{0\}$ and $M\times[0,1]$ together.
We also denote $f$ (resp.~$ f'$) as a $d$-simplex on the orientation reversing wall contained in $N\times\{0\}$ (resp.~$M\times[0, 1]$) respectively,  which shares $e$ (resp.~$e'$).
Then, we choose the assignment of $\vartheta_e, \vartheta_{e'}$, such that (see Fig.~\ref{fig:attach} (b')) 
\begin{itemize}

\item we place $\theta_{e'}$, $\overline{\theta}_{e'}$  on $e'$ such that the color of the Grassmann variable on $e'$ on one side of the orientation reversing wall coincides with that of $f$.
\item we place $\theta_e, \overline{\theta}_e$ on $e$ such that the color of the Grassmann variable on $e$ on one side of the orientation reversing wall differs from that of $e'$.
\end{itemize}

We emphasize that such redefinition or a specific choice of assignment does not affect the quadratic property~\eqref{eq:Quadboundary}.
After these preparations, to see~\eqref{eq:retriangle}, we first observe the quadratic property of $\tilde{\sigma}(\alpha;\beta):=\sigma(N\times\{0\})\sigma(M\times[0,1])\sigma^{-1}(N\times\{1\})$,
\begin{equation}
\tilde{\sigma}(\alpha;\beta)\tilde{\sigma}(\alpha';\beta')=\tilde{\sigma}(\alpha+\alpha';\beta+\beta')(-1)^{\int_{\partial K}\beta\cup_{d-1}\beta'},
\label{eq:Quadtilde}
\end{equation}
which can be seen by applying quadratic property of $\sigma$~\eqref{eq:Quadboundary} on $N\times\{0\}$, $M\times[0,1]$, $N\times\{1\}$.
Note that~\eqref{eq:Quadtilde} is satisfied for
\begin{equation}
    \tilde{\sigma}'(\alpha;\beta)=(-1)^{\int_{K}\Sq^2(\beta)},
\end{equation}
where we set $\Sq^2(\beta):=\beta\cup_{d-2}\beta+\delta\beta\cup_{d-1}\beta$. Thus, we can express $\tilde{\sigma}(\alpha;\beta)$ as $(-1)^{\int_K \Sq^2(\beta)}$ up to linear term,
\begin{equation}
   \tilde{\sigma}(\alpha;\beta)=(-1)^{\int_K \Sq^2(\beta)}(-1)^{\sum_{f\in K}\chi(f)\beta(f)}. 
\end{equation}
The linear term is fixed by computing $\tilde{\sigma}(\alpha;\beta)$ explicitly in the simplest case; 
$\beta=\delta\lambda$ on $\partial K$, and $\lambda(e)=1$ on a single $(d-1)$-simplex of $\partial K$, otherwise 0.
If we take a barycentric subdivision on $\partial K$, we can see that $\tilde{\sigma}(\alpha=0;\delta\lambda)=-1$ when $\lambda$ is nonzero on the dual of $w_2+w_1^2$ described in Sec.~\ref{sec:retri}, at least if nonzero $\lambda$ is away from the boundary of $N\times\{0\}$, $M\times[0,1]$, $N\times\{1\}$. When $\lambda$ is nonzero on the boundary, we should be more careful. For instance, let $\lambda$ be nonzero on $M\times\{0\}$. 
First, we discuss the case when $\lambda$ is nonzero away from the orientation reversing wall of $M$. Thanks to the above redefinition of $\sigma(M\times[0,1])$, we can see that $\sigma(N\times\{0\})$ and $\sigma(M\times[0,1])$ have the opposite sign. Hence, we have $\tilde{\sigma}(\alpha;\beta)=-1$. 
Next, let us examine the case when $\lambda(e)=1$ on the orientation reversing wall. Let $f$, $f'$ be two $d$-simplices on the side of $e$ in the wall, which is contained in $N\times\{0\}$, $M\times[0,1]$, respectively. Then, one can see that $\tilde{\sigma}(\lambda;\delta\lambda)=-1$, if the Grassmann variables on $f, f'$ on one side of the wall have the same color, otherwise $+1$. 
Thus, now we can say that $\tilde{\sigma}(\lambda;\delta\lambda)=-1$ when $\lambda(e)$ is nonzero on the dual of $w_2+w_1^2$ of $\partial K$, even when $e$ lies in $M\times \{0\}$.
Thus, we can fix the linear term as~\eqref{eq:retriangle} up to linear and gauge invariant term.

Next, we demonstrate~\eqref{eq:M01}. 
Note that~\eqref{eq:Quadboundary} is satisfied for
\begin{align}
    \sigma'(M\times[0,1])=(-1)^{\int_{M\times[0,1]} \Sq^2(\alpha)}.
\end{align}
Thus, we can express $\sigma(M\times[0,1])$ as $(-1)^{\int_{M\times[0,1]} \Sq^2(\alpha)}$ up to linear term,
\begin{equation}
   \sigma(M\times[0,1])=(-1)^{\int_{M\times[0,1]} \Sq^2(\alpha)}(-1)^{\sum_{e\in M\times[0,1]}\lambda(e)\alpha(e)}. 
\end{equation}
Note  that we are assuming that $\alpha$ extends to $M\times[0,1]$.
The linear term is again fixed by computing $\sigma(M\times[0,1])$ explicitly in the simplest case; $\alpha(e)=1$ on a single $(d-1)$-simplex, otherwise 0. 
If we take a barycentric subdivision on $M\times[0,1]$, we can see that $\tilde{\sigma}(\alpha;\delta\alpha)=-1$ when $\lambda$ is nonzero on the dual of $w_2+w_1^2$ described in Sec.~\ref{sec:retri}, at least if $\alpha$ is nonzero away from the boundary of  $M\times[0,1]$. When $\alpha$ is nonzero on the boundary, it requires more careful treatment. 
In this situation, by arranging the sign of $f_0$ chosen to be identical to $t$, we can see that $\sigma(\alpha;\delta\alpha)=-1$ when $\alpha$ is nonzero in $M\times\{0\}$ away from the orientation reversing wall.
Next, let us examine the case that $\alpha(e')$ is nonzero for $e'$ contained in the orientation reversing wall. Let $f$, $f'$ be two $d$-simplices on the side of $e'$ in the wall, which is contained in $N\times\{0\}$, $M\times[0,1]$, respectively. Then, one can see that $\tilde{\sigma}(\alpha;\delta\alpha)=-1$, if the Grassmann variables on $f, f'$ on one side of the wall have the same color, otherwise $+1$, thanks to the choice of assignment of Grassmann variables in $M\times\{0\}$. 
Thus, now we can say that $\tilde{\sigma}(\alpha;\delta\alpha)=-1$ when $\alpha(e)$ is nonzero on the dual of $w_2+w_1^2$ of $\partial K$, even when $e$ lies in $M\times \{0\}$. Thus, we have the fixed the linear term as~\eqref{eq:M01}. 

Combining~\eqref{eq:retriangle} with~\eqref{eq:M01}, the variation of $\sigma(N)$ under re-triangulation and gauge transformation is given by
\begin{equation}
    (-1)^{\int_{M\times[0,1]}(\Sq^2(\alpha)+(w_2+w_1^2)\cup \alpha)+\int_{N\times[0,1]}(\Sq^2(\beta)+(w_2+w_1^2)\cup \beta)}.
\end{equation}
On the other hand, the variation of $(-1)^{\int_M\eta\cup\alpha+\int_N\eta\cup\beta}$ is given by
\begin{equation}
    (-1)^{\int_{M\times[0,1]}(w_2+w_1^2)\cup \alpha+\int_{N\times[0,1]}(w_2+w_1^2)\cup \beta},
\end{equation}
where $\eta$ specifies a pin$_-$ structure.
Hence, the variation of the Grassmann integral $z[\eta; \alpha,\beta]=\sigma(\alpha;\beta)(-1)^{\int_M\eta\cup\alpha+\int_N\eta\cup\beta}$ becomes
\begin{equation}
    (-1)^{\int_{M\times[0,1]}\Sq^2(\alpha) +\int_{N\times[0,1]}\Sq^2(\beta)}.
    \label{gwv}
\end{equation}
In the pin$_+$ case, the variation of $z[\eta; \alpha,\beta]$ is instead given by
\begin{equation}
    (-1)^{\int_{M\times[0,1]}
    (\Sq^2(\alpha)+w_1^2\cup\alpha) +\int_{N\times[0,1]}
    (\Sq^2(\beta)+w_1^2\cup\beta)}.
    \label{gwvp}
\end{equation}

\subsection{Gapped boundary for the Gu-Wen pin phase}
After all these preparations, it is a simple matter to show that the boundary gauge theory \eqref{gwb} correctly couples to the bulk Gu-Wen pin SPT phase.
Indeed, the partition function of the coupled system has the action \begin{equation}
z[\eta;\alpha,\beta] (-1)^{-\int_M h^*x_d + \int_N g^*y_{d+1}} \label{combined}
\end{equation}
for both pin$_-$ and pin$_+$ case, where we take $\alpha=h^*m_{d-1}$ and $\beta=g^*n_d$.
The first term in \eqref{combined} has the variation \eqref{gwv} (resp.~\eqref{gwvp}) in pin$_-$ (resp.~pin$_+$) case,
whereas the second term in \eqref{combined} has the variation 
\begin{equation}
(-1)^{\int_{M\times[0,1]} (h^*\delta_{\rho} x_d-g^*y_{d+1}) -\int_{N\times[0,1]} g^*\delta_{\rho} y_{d+1} }.
\end{equation}
These two variations cancel since we have  $\delta_{\rho} y_{d+1}=\Sq^2(n_d)$ (resp.~$\delta_{\rho} y_{d+1}=\Sq^2(n_d)+\rho_1^2\cup n_d$) and $y_{d+1}$ pulls back to $\Sq^2(m_{d-1})+\delta_{\rho} x_d$ (resp.~$\Sq^2(m_{d-1})+(p^*\rho_1)^2\cup m_{d-1}+\delta_{\rho} x_d$) in pin$_-$ (resp.~pin$_+$) case.
This is what we wanted to achieve.

\section{Time reversal anomaly of (2+1)d pin$_+$ TQFT}
\label{sec:anomaly}
In this final section, 
we apply our construction to the analysis of (2+1)d time reversal anomaly of class DIII, which is classified by $\Omega_4^{\mathrm{pin}_+}(\mathrm{pt})=\bZ_{16}$~\cite{KirbyTaylor}. 

In~\cite{Fidkowski2014}, the authors provided (presumably a bosonic shadow of) an anomalous theory based on the
(3+1)d Walker-Wang model~\cite{WalkerWang2011}. 
Their (3+1)d Walker-Wang model is constructed from a data of (2+1)d TQFT characterized by a premodular braided fusion category equipped with a transparent fermion, whose line operator generates a $\bZ_2$ 1-form symmetry.
By construction, the resulting (3+1)d Walker-Wang model admits a gapped boundary described by the given (2+1)d TQFT. 
Later,~\cite{Wang2017indicator} conjectured the indicator formula that determines the $\bZ_{16}$-valued (2+1)d pin$_+$ anomaly, from the data of (2+1)d TQFT realized on a boundary of the (3+1)d Walker-Wang model. The conjectured formula was demonstrated in~\cite{Tachikawa2017more}, based on the argument that prepares the Hilbert space of pin$_+$TQFT on a boundary of a non-orientable manifold.

The above background motivates us to revisit the indicator formula of the time-reversal anomaly in (2+1)d, by coupling the Walker-Wang model with the Grassmann integral we have constructed above. 
We aim to obtain the indicator formula for the $\bZ_{16}$ anomaly, in terms of the data of  the shadow TQFT.
Suppose we have constructed the shadow of a (3+1)d pin$_+$ SPT phase, described by a (3+1)d Walker-Wang model. Then, the Walker-Wang model is equipped with a line operator $f$, which generates an anomalous $\bZ_2$ 2-form symmetry characterized by
\begin{equation}
(-1)^{\int (\Sq^2(\alpha)+w_1^2\cup\alpha)},
\end{equation}
where $\alpha$ is the background gauge field. 
Then, the invertible pin$_+$ theory for an SPT phase is given by coupling with the Grassmann integral as
\begin{equation}
Z[M, \eta]=\sum_{[\alpha]\in H^3(M, \bZ_2)}Z_{\mathrm{WW}}[M, \alpha]\cdot \sigma(M, \alpha) (-1)^{\int_M \eta\cup\alpha},
\label{pin4part}
\end{equation}
where $Z_{\mathrm{WW}}[M,\alpha]$ denotes the partition function of the Walker-Wang model in the presence of the background 3-form gauge field. $\eta$ specifies a pin$_+$ structure that satisfies $\delta\eta=w_2$. 
Since $\Omega_4^{\mathrm{pin}_+}(\mathrm{pt})=\bZ_{16}$ is generated by $\mathbb{RP}^4$ equipped with a pin$_+$ structure, one should be able to construct the indicator formula by evaluating~\eqref{pin4part} for $M=\mathbb{RP}^4$. In this case, we sum over $[\alpha]\in H^3(\mathbb{RP}^4, \bZ_2)=\bZ_2$, where a nontrivial element of $Z^3(\mathbb{RP}^4, \bZ_2)$ corresponds to the insertion of a single line operator $f$ along a homotopically nontrivial line of $\mathbb{RP}^4$. The Grassmann integral $i^{Q_{\eta}[\alpha]}=\sigma(M,\alpha)(-1)^{\int_M \eta\cup\alpha}$ is again computed via the quadratic property~\eqref{eq:unorientedquad},
\begin{equation}
Q_{\eta}[\alpha]+Q_{\eta}[\alpha']=Q_{\eta}[\alpha+\alpha']+2\int_{M}\alpha\cup_{2}\alpha'.
\end{equation}
When $\alpha$ is a
nontrivial element of $Z^3(\mathbb{RP}^4, \bZ_2)$, one can see that
\begin{equation}
2\int_{\mathbb{RP}^4}\alpha\cup_2\alpha=2 \quad\mod 4.
\end{equation}
Thus, we can show that $Q_{\eta}[\alpha]=\pm 1$ mod 4, for a nontrivial $\alpha$. Such two choices of $Q_{\eta}[\alpha]$ correspond to different choice of pin$_+$ structure. Hence, the indicator formula has the form of 
\begin{equation}
Z[\mathbb{RP}^4]=Z_{\mathrm{WW}}[\mathbb{RP}^4, 0]\pm i\cdot Z_{\mathrm{WW}}[\mathbb{RP}^4, \alpha],
\end{equation}
for nontrivial $\alpha\in Z^3(\mathbb{RP}^4, \bZ_2)$. Fortunately, the partition function of the Walker-Wang model on $\mathbb{RP}^4$ was explicitly computed in~\cite{Barkeshli2016}.
The result is expressed via the data of (2+1)d TQFT on boundary,
\begin{equation}
Z_{\mathrm{WW}}[\mathbb{RP}^4, 0]=\frac{1}{\mathcal{D}}\sum_{\overline{p}={\sf R}(p)}\eta_pd_p e^{i\theta_p},
\label{RP4part}
\end{equation}
where $d_p$ is quantum dimension of $p$, $\mathcal{D}$ is total dimension characterized by $\mathcal{D}^2:=\sum_pd_p^2$, and $\theta_p$ is $\mathbb{R}/2\pi\bZ$-valued topological spin of $p$. $\sf R$ denotes an orientation reserving symmetry, and
$\eta_p$ is a quantity that characterizes the symmetry fractionalization of an anyon $p$;
$\eta_p$ is defined as the $\sfR$ eigenvalue of the $\sfR$ symmetric state $\ket{p, \sfR(p)}$ constructed on the Hilbert space on $S^2=\partial D^3$, 
where $\sfR$ is implemented as the antipodal map of $D^3$, and two anyons $p$, $\sfR(p)$ are located on $S^2$ in an $\sfR$ symmetric (i.e., antipodal) fashion. Namely, we have
\begin{equation}
\sfR\ket{p, \sfR(p)}=\ket{\sfR^2(p), \sfR(p)}=:\eta_p\cdot\ket{p, \sfR(p)}.
\label{defeta}
\end{equation}
In the first equation in~\eqref{defeta}, we note that $\sfR$ permutes the position of two quasiparticles.  The state $\ket{p, \sfR(p)}$ exists only when 
$p, \sfR(p)$ fuse into vacuum; $\overline{p}=\sfR(p)$, otherwise $\eta_p$ becomes ill-defined.
Accordingly, the summation runs over quasiparticles such that $\bar{p}=\sfR(p)$ in~\eqref{RP4part}. 

By imitating the logic in~\cite{Barkeshli2016}, we can also compute $Z_{\mathrm{WW}}[\mathbb{RP}^4, \alpha]$ for nontrivial background gauge field $\alpha$. In such a situation, the background field is realized as a single insertion of the transparent line operator $f$, along a homotopically nontrivial loop in $\mathbb{RP}^4$. As we examine in detail in Appendix~\ref{app:ww}, the single insertion of an $f$ line amounts to evaluating the symmetry fractionalization on the Hilbert space on $S^2=\partial D^3$, in the presence of a single $f$ particle at the center of $D^3$; concretely,  we prepare a $\sfR$ symmetric state $\ket{p, \sfR(p)}$ constructed on the Hilbert space on $S^2=\partial D^3$, where $p$ and $\sfR(p)$ lines fuse into an $f$ particle. The corresponding $f$ line ends at the center of $D^3$. 
If we also denote the $\sfR$ eigenvalue of such state $\ket{p, \sfR(p)}$ as $\eta_p$, the partition function in the presence of nontrivial background field is expressed as
\begin{equation}
Z_{\mathrm{WW}}[\mathbb{RP}^4, \alpha]=\frac{1}{\mathcal{D}}\sum_{\overline{p}=f\cdot{\sf R}(p)}\eta_pd_p e^{i\theta_p},
\label{RP4partf}
\end{equation}
where we sum over $p$ such that $p, \sfR(p)$ fuse into $f$. After all, the indicator is expressed as
\begin{equation}
Z[\mathbb{RP}^4]=\frac{1}{\mathcal{D}}\left(\sum_{\overline{p}={\sf R}(p)}\eta_pd_p e^{i\theta_p}\pm i\cdot\sum_{\overline{p}=f\cdot{\sf R}(p)}\eta_pd_p e^{i\theta_p}\right),
\end{equation}
which reproduces the indicator formula proposed in~\cite{Wang2017indicator}, if we identify the above definition of $\eta_p$ as $\mathcal{T}_p^2$ in~\cite{Wang2017indicator}.
\footnote{Our definition of the total dimension $\mathcal{D}$ is related to that of~\cite{Wang2017indicator} by $\mathcal{D}=\sqrt{2}D$, since our total dimension $\mathcal{D}$ counts the contribution of the transparent particle $f$, while $D$ in~\cite{Wang2017indicator} does not.}
 The validity of such identification should be demonstrated for explicit lattice models, which is left for future work.

\section*{Acknowledgements}
The author thanks Shinsei Ryu and Yuji Tachikawa for reading the draft of this paper and giving helpful suggestions and improvement. The author is grateful to Kantaro Ohmori, Shinsei Ryu, and Yuji Tachikawa for useful discussions.
The author also acknowledges the hospitality of Kadanoff Center for Theoretical Physics.
The author is supported by Advanced Leading
Graduate Course for Photon Science (ALPS) of Japan Society for the Promotion of Science (JSPS). 

\appendix
\section{Partition function of the (3+1)d Walker-Wang model}
\label{app:ww}
In this appendix, following the logic of~\cite{Barkeshli2016}, we compute the partition function $Z_{\mathrm{WW}}[\mathbb{RP}^4, \alpha]$ of the Walker-Wang model on $\mathbb{RP}^4$, with or without background gauge field.

\subsection{Gluing relation}
We compute the partition function on a 4-manifold by decomposing the manifold into simpler manifolds for which partition functions are easier to evaluate
and computing the partition function part by part. This procedure is performed via applying the gluing relation for the path integral.
Here, let us review the application of the gluing relation to the Walker-Wang model,
which is required for explicit computations, following Ref.~\cite{Barkeshli2016, Walker06}.

The data of (2+1)d TQFT (a braided fusion category $\mathcal{B}$) defines a (3+1)d TQFT known as the Walker-Wang model.
To consider the path integral of the Walker-Wang model on a 4d manifold $M^4$, 
we first specify the configuration of fields on the boundary $c\in\calC(\p M^4)$, where $\calC(\p M^4)$ denotes a set of boundary conditions. 

Here, the set of boundary conditions on a 3-manifold $M^3$, $\calC(M^{3})$, is defined as the set of all configurations of anyon diagrams on $M^{3}$, based on the braided fusion category $\mathcal{B}$.
If $M^3$ has boundary, we denote $\calC(M^{3}; c)$ as the configuration space of anyon diagrams on $M^{3}$, under the boundary condition $c$ on $\p M^3$.

Then, the Hilbert space $\calV(M^{3}; c)$ of the bulk-boundary coupled system is defined as the formal linear superposition of anyon diagrams $\mathbb{C}[\calC(M^{3}; c)]$, modded out by equivalence relations (e.g., fusions, $F$ and $R$ moves in $\mathcal{B}$),
\begin{align}
\calV(M^{3}; c):=\mathbb{C}[\calC(M^{3}; c)]/\sim.
\end{align}
Then, the path integral $Z(M^{4})$ is a map from $\calV(\p M^{4})$ to a number,
\begin{align}
Z(M^{4}):\quad \calV(\p M^{4})\mapsto \mathbb{C}.
\end{align}
We will write the value as $Z(M^{4})[c]$, for $c\in\calV(\p M^{4})$. 
The inner product in $\calV(M^{3}; c)$ is defined via the bulk partition function as
\begin{align}
\langle x|y\rangle_{\calV(M^{3}; c)}:=Z(M^{3}\times I)[\overline{x}\cup y],
\end{align}
where $M^{3}\times I$ is a $4$-manifold pinched at $\partial M^3\times I$ by the identification $(b, s)\sim(b, t)$ for $b\in\partial M^3$ and $s,t\in I$, 
so that $\partial(M^{3}\times I)=M^3\cup\overline{M}^3$. Furthermore, $\overline{x}$, $y$ specifies boundary conditions on $\overline{M}^3$, $M^3$ respectively, where
$\overline{x}$ denotes the field configuration on $\overline{M}^3$ given by reversing the orientation of $x$. 

\begin{figure}[htb]
\centering
\includegraphics{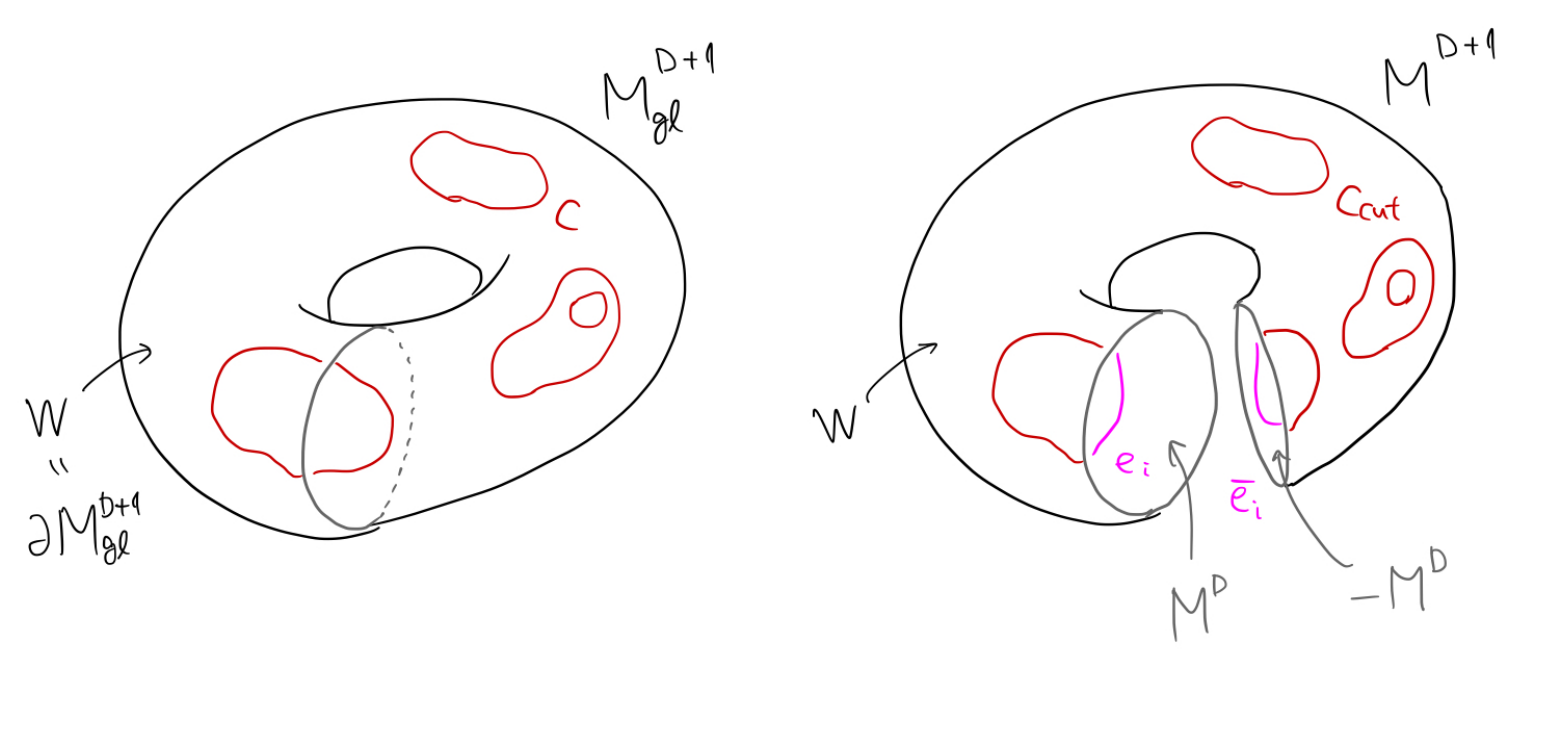}
\caption{Illustration of the gluing relation.}
\label{fig:gluing}
\end{figure}

Now, let us describe the gluing relation.
Let $M^{4}$ be a $4$-manifold whose boundary is $\p M^{4}=M^{3}\cup \overline{M}^{3}\cup W$, 
and $M^{4}_{\mathrm{gl}}$ be a $4$-manifold which is given by gluing the boundary of $M^{4}$ along $M^{3}$ and $\overline{M}^{3}$.
Then, the partition function $Z(M^{4}_{\mathrm{gl}})[c]$ on $M^{4}_{\mathrm{gl}}$ 
with a boundary condition $c\in \calV(W)$ on $W=\partial{M^{4}_{\mathrm{gl}}}$ is evaluated via the following gluing relation,
\begin{align}
Z(M^{4}_{\mathrm{gl}})[c]=\sum_{e_i}\frac{Z(M^{4})[c_{\text{cut}}\cup e_i\cup \overline{e}_i]}{\langle e_i|e_i \rangle_{\calV(M^{3}; c_{\mathrm{cut}}^{2})}},
\label{gluing}
\end{align}
where $c_{\text{cut}}$ is the boundary condition inherited from $c$ after the cut, and $c_{\mathrm{cut}}^{2}$ is the restriction of $c_{\text{cut}}$ to $\p M^{3}$.
We denote an orthonormal basis of $\calV(M^{3}; c_{\text{cut}}^{2})$ as $\{e_i\}$.
We illustrate the situation of the gluing relation in Fig.~\ref{fig:gluing}.

\subsection{Handle decomposition of $\bRP^4$}
Let us turn to the explicit computations of $Z(\mathbb{RP}^4)$.
We can compute the partition function on $\bRP^4$ via gluing relations, by decomposing $\bRP^4$ into simpler manifolds for which partition functions are easier to evaluate.
To do this, we employ handle decomposition on $\bRP^4$, which takes $\bRP^4$ apart into 4-balls.

For $1\le k\le d$, a $k$-handle in $d$ dimension is defined as a pair $(D^k\times D^{d-k}, S^{k-1}\times D^{d-k})$. 
$S^{k-1}\times D^{d-k}\subset\p{(D^k\times D^{d-k})}$ is called an attaching region of the $k$-handle. The 0-handle is defined as $D^d$.
We think of attaching a $k$-handle to a $d$-manifold $M_0$ with boundary, by an embedding of attaching region 
$\phi: S^{k-1}\times D^{d-k}\mapsto\p M_0$ such that the image of $\phi$ is contained in $\p M_0$.
It is known that every compact $d$-manifold $M$ without boundary allows a handle decomposition, i.e.,
$M$ is developed from a 0-handle by successively attaching to it handles of dimension $d$.

We can see that $\bRP^4$ is composed of single $k$-handles for each $k=0, 1, 2, 3, 4$ by the following steps.

\begin{enumerate}
\item To see this, it is convenient to think of $\bRP^4$ as $D^4$ 
with its boundary $\p D^4=S^3$ identified by an antipodal map. 
First, we begin with locating a small 0-handle containing the center of $D^4$.

\item Next, we attach a 1-handle $(D^1\times D^{3}, S^{0}\times D^{3})$ to the 0-handle. 
The attaching region of a 1-handle consists of two 3-balls, $S^{0}\times D^{3}=D^3\cup D^3$.
We attach one of these $D^3$s to the boundary $S^3$ of 0-handle, by identifying with a small $D^3$ in $S^3$.
Then, we radially extend a 1-handle from the attached $D^3$, which tunnels through the antipodal map and returns to the 0-handle again.
Eventually, we attach the other $D^3$ of the 1-handle to the 0-handle.
We denote the composition of the $0, 1, \dots k$ handles in $\bRP^4$ as $\bRP^4_k$. At this point, we have constructed $\bRP^4_1$.

\item Then, we attach a 2-handle $(D^2\times D^{2}, S^{1}\times D^{2})$ to $\bRP^4_1$. 
We note that $\bRP^4_1=(D^3\times S^1)/\sigma$, where $\sigma$ is the $\bZ_2$ action on $D^3\times S^1$ defined as the composite of antipodal maps.
The attaching region $D^2\times S^{1}$ is embedded in $\p(\bRP^4_1)=(S^2\times S^1)/\sigma$, via embedding a small $D^2$ in $S^2$.

\item Likewise, we attach a 3-handle $(D^3\times D^{1}, S^{2}\times D^{1})$ to $\bRP^4_2=(D^2\times S^2)/\sigma$
by embedding the attaching region in $ \p(\bRP^4_2)=(S^1\times S^2)/\sigma$,
\footnote{We note the abuse of notation; $\sigma$ always denotes the composite of antipodal maps in this context.} via embedding a small $D^1$ in $S^1$. 

\item Finally, we complete $\bRP^4$ with attaching a 4-handle $(D^4, S^{3})$ to $\bRP^4_3=(D^1\times S^3)/\sigma$, 
by identifying the attaching region with $\p(\bRP^4_3)=(S^0\times S^3)/\sigma=S^3$. 
\end{enumerate}

\subsection{Partition function on $\bRP^4$}
Now we can compute $Z(\bRP^4)[\alpha]$ by successively applying gluing relations in each process of the handle decomposition. In the presence of nontrivial background gauge field $\alpha\in Z^3(\bRP^4, \bZ_2)$, $\alpha$ amounts to inserting a single Wilson line $l_f$ of $f$, along a loop in $\bRP^4$ that intersects the crosscap of $\bRP^4$ once. Here, we choose the configuration of $l_f$, such that $l_f$ is contained in $\bRP^4_1=(D^3\times S^1)/\sigma$.  $l_f$ runs in the $S^1$ direction of $(D^3\times S^1)/\sigma$, living at the center of $D^3$ of $(D^3\times S^1)/\sigma$. We denote the partition function on $M$ in the presence of such a line operator, simply as $Z(M; l_f)$.

Then, the computation of $Z(\bRP^4; l_f)$ proceeds as follows.
\begin{enumerate}
\item First, we decompose $\bRP^4$ into $\bRP^4_3$ and a 4-handle, along the attaching region $S^3$. 
Since there is no nontrivial anyon diagram on $S^3$ up to equivalence relations, 
only the empty diagram $\varnothing$ contributes to the boundary condition. Hence, the gluing relation becomes
\begin{align}
Z(\bRP^4)=\frac{Z(\bRP^4_3; l_f)[\varnothing] Z(D^4)[\varnothing]}{\langle\varnothing|\varnothing\rangle_{\calV(S^3)}}.
\end{align}
As shown in~\cite{Barkeshli2016}, we can see that $Z(D^4)[\varnothing]=1/\calD$ and $\langle\varnothing|\varnothing\rangle_{\calV(S^3)}=Z(S^3\times D^1)[\varnothing]=1/\calD^2$, where $\calD$ is the total dimension of anyons. Thus, 
\begin{align}
Z(\bRP^4)=\frac{Z(\bRP^4_3; l_f)[\varnothing] \cdot 1/\calD}{1/\calD^2}=\calD\cdot Z(\bRP^4_3; l_f)[\varnothing].
\label{RP4RP43}
\end{align}

\item Next, we decompose $\bRP^4_3$ into $\bRP^4_2$ and a 3-handle, along the attaching region $S^2\times D^1$. 
Similarly, only the empty diagram $\varnothing$ contributes to the bounndary condition on the cut $S^2\times D^1$. Gluing relation becomes
\begin{align}
Z(\bRP^4_3; l_f)[\varnothing]=\frac{Z(\bRP^4_2; l_f)[\varnothing] Z(D^4)[\varnothing]}{\langle\varnothing|\varnothing\rangle_{\calV(S^2\times D^1; \varnothing)}}.
\end{align}
As shown in~\cite{Barkeshli2016}, we have $Z(D^4)[\varnothing]=1/\calD$ and $\langle\varnothing|\varnothing\rangle_{\calV(S^2\times D^1; \varnothing)}=Z(S^2\times D^2)[\varnothing]=1$. Thus,
\begin{align}
Z(\bRP^4_3; l_f)[\varnothing]=1/\calD\cdot Z(\bRP^4_2; l_f)[\varnothing].
\end{align}
Combining this expression with~\eqref{RP4RP43}, we have
\begin{align}
Z(\bRP^4; l_f)=Z(\bRP^4_2; l_f)[\varnothing].
\label{RP4RP42}
\end{align}

\item Then, we decompose $\bRP^4_2$ into $\bRP^4_1$ and a 2-handle, along the attaching region $S^1\times D^2$.
The boundary condition on the cut $S^1\times D^2$ is labeled by the loop $l_a$ of anyon $a$ going around the $S^1$.
Gluing relation becomes
\begin{align}
Z(\bRP^4_2; l_f)[\varnothing]=\sum_a\frac{Z(\bRP^4_1; l_f)[l_a^{(+1)}] Z(D^4)[l_a]}{\langle l_a| l_a\rangle_{\calV(S^1\times D^2; \varnothing)}}.
\end{align}
Here, we have an $l_a$ line on $\p(\bRP^4_1)=(S^2\times S^1)/\sigma$ going along $(\{p\}\times S^1)/\sigma$, where $p$ denotes some point of $S^2$.
The notation $l_a^{(+1)}$ means that the $l_a$ diagram has $+1$ framing, as demonstrated in~\cite{Barkeshli2016}. 
For $Z(D^4)[l_a]$, we have a bubble of $l_a$ loop on $S^3=\p D^4$ weighted by quantum dimension $d_a$. 
Hence, $Z(D^4)[l_a]=d_a Z(D^4)[\varnothing]$.
As shown in~\cite{Barkeshli2016}, we have $Z(D^4)[\varnothing]=1/\calD$, $\langle l_a| l_a\rangle_{\calV(S^1\times D^2; \varnothing)}=1$. Therefore,
\begin{align}
Z(\bRP^4_2; l_f)[\varnothing]=\frac{1}{\calD}\sum_{a}d_a\cdot Z(\bRP^4_1; l_f)[l_a^{(+1)}].
\label{RP42RP41}
\end{align}

\item Finally, we evaluate $Z(\bRP^4_1; l_f)[l_a^{(+1)}]$. 
$\bRP^4_1=(D^3\times S^1)/\sigma$ is a twisted solid torus $D^3\rtimes S^1$,
where the twist is defined as an antipodal map of $D^3$.

Let us recall the configuration of line operators for $Z(\bRP^4_1; l_f)[l_a^{(+1)}]$.
First, we have an $l_a$ line on the boundary. The $l_a$ line on the boundary $S^2\rtimes S^1$ of $D^3\rtimes S^1$ looks like a worldline of a pair of anyon $a, \sfR(a)$ living in north and south pole of $S^2$ respectively, which is identified with each other at the twist. (Here, note that the antipodal map acts on anyon label as $a\to\sfR(a)$.)

Moreover, we have an $l_f$ line in the bulk living at the center of $D^3$.
To apply the gluing relation, we cut $\bRP^4_1=D^3\rtimes S^1$ at a point of $S^1$ where we twist $D^3$. On the cut section $D^3$, we have a pair of anyons $a, \sfR(a)$ located in the antipodal fashion on the boundary, and also a single $f$ particle at the center of $D^3$, see Fig.~\ref{fig:antipodal}. 
We write the Hilbert space on the cut with such a configuration of anyons as $\calV(D^3; a, \sfR(a), f)$. We note that such a state exist iff $a$ and $\sfR(a)$ fuse into $f$. 
Then, the boundary condition on the section $D^3$ is a diagram $e$ which joins $a, \sfR(a)$ and $f$ together.

Now, recall that we have defined $\eta_a$ as the eigenvalue of the antipodal map on the state in $\calV(D^3; a, \sfR(a), f)$.
Since we operate the antipodal map when gluing the cut section $D^3$, it picks up the eigenvalue $\eta_a$ by acting on the state of the section.

After all, when $\bar{a}=f\cdot\sfR(a)$, the gluing relation becomes
\begin{align}
\begin{split}
Z(D^3\rtimes S^1; l_f)[l_a^{(+1)}]&=e^{i\theta_a}\cdot\eta_a\cdot\frac{Z(D^3\times I)[e \cup\overline{e}]}{\langle e|e\rangle_{\calV(D^3; a, \sfR(a), f)}} \\
&=e^{i\theta_a}\cdot\eta_a,
\label{eq:RP41}
\end{split}
\end{align}
otherwise we have $Z(D^3\rtimes S^1, l_f)[l_a^{(+1)}]=0$.
Here, the framing $+1$ contributes as topological spin $e^{i\theta_a}$ of $a$.

\begin{figure}[htb]
\centering
\includegraphics{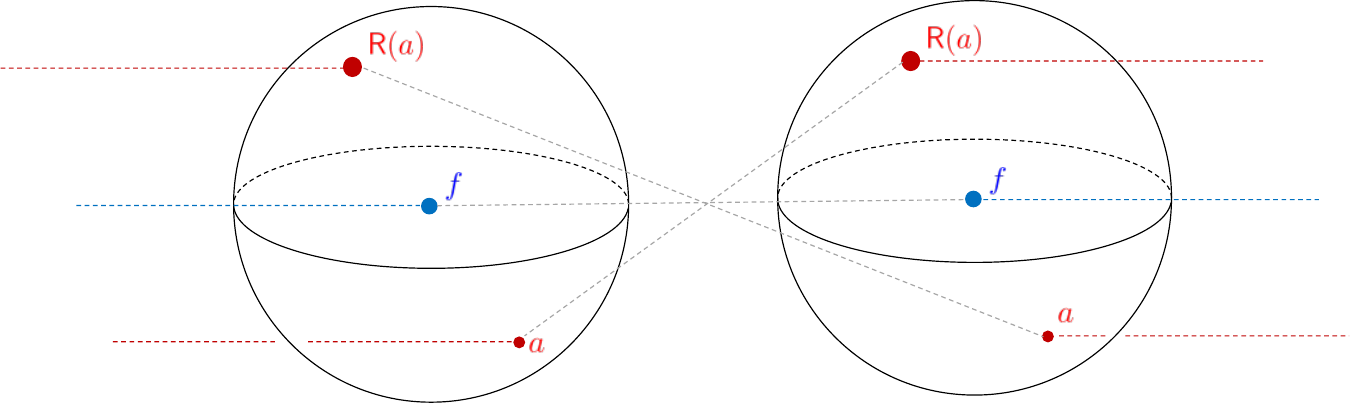}
\caption{Configuration of anyons on the cut section $D^3$. The antipodal map of $D^3$ acts on the state with particles $a, \sfR(a), f$.}
\label{fig:antipodal}
\end{figure}

\end{enumerate}

Combining~\eqref{RP4RP42}, \eqref{RP42RP41} with~\eqref{eq:RP41}, we eventually obtain the partition function as
\begin{align}
Z(\mathbb{RP}^4; l_f)&=\frac{1}{\mathcal{D}}\sum_{\bar{a}=f\cdot\sfR(a)}d_a\eta_ae^{i\theta_a}.
\label{RP4parti}
\end{align}
On the other hand, when the background gauge field is trivial, we just have to discard the $f$ line in the argument described above, and we have
\begin{align}
Z(\mathbb{RP}^4; 0)&=\frac{1}{\mathcal{D}}\sum_{\bar{a}=\sfR(a)}d_a\eta_ae^{i\theta_a}.
\label{RP4parti0}
\end{align}

\bibliographystyle{ytphys}
\baselineskip=.95\baselineskip
\bibliography{ref}

\end{document}